\documentclass[twocolumn,prc,superscriptaddress,unsortedaddress,showpacs,preprintnumbers,amsmath,amssymb]{revtex4}
\usepackage{tipa}

\usepackage{graphicx}
\usepackage{dcolumn}
\usepackage{bm}
\usepackage{color}
\UseRawInputEncoding

\def\beq{\begin{equation}}
\def\eeq{\end{equation}}
\def\bea{\begin{eqnarray}}
\def\eea{\end{eqnarray}}

\def\fun#1#2{\lower3.6pt\vbox{\baselineskip0pt\lineskip.9pt
  \ialign{$\mathsurround=0pt#1\hfil##\hfil$\crcr#2\crcr\sim\crcr}}}

\preprint{}

\begin{document}

\title{Exact solution of
the Brueckner-Bethe-Goldstone equation with three-body forces in nuclear matter}

\author{X. -L. Shang}\email[ ]{shangxinle@impcas.ac.cn}
 \affiliation{Institute
of Modern Physics, Chinese Academy of Sciences, Lanzhou 730000,
China}\affiliation{School of Nuclear Science and Technology,
University of Chinese Academy of Sciences, Beijing 100049, China}
\affiliation{CAS Key Laboratory of High Precision Nuclear
Spectroscopy, Institute of Modern Physics, Chinese Academy of
Sciences, Lanzhou 730000, China}
\author{J.-M. Dong}
 \affiliation{Institute
of Modern Physics, Chinese Academy of Sciences, Lanzhou 730000,
China}\affiliation{School of Nuclear Science and Technology,
University of Chinese Academy of Sciences, Beijing 100049,
China}\affiliation{CAS Key Laboratory of High Precision Nuclear
Spectroscopy, Institute of Modern Physics, Chinese Academy of
Sciences, Lanzhou 730000, China}

\author{W. Zuo}
\affiliation{Institute of Modern Physics, Chinese Academy of
Sciences, Lanzhou 730000, China}\affiliation{School of Nuclear
Science and Technology, University of Chinese Academy of Sciences,
Beijing 100049, China}\affiliation{CAS Key Laboratory of High
Precision Nuclear Spectroscopy, Institute of Modern Physics,
Chinese Academy of Sciences, Lanzhou 730000, China}

\author{P. Yin}
\affiliation{Institute of Modern Physics, Chinese Academy of
Sciences, Lanzhou 730000, China}\affiliation{School of Nuclear
Science and Technology, University of Chinese Academy of Sciences,
Beijing 100049, China}\affiliation{CAS Key Laboratory of High
Precision Nuclear Spectroscopy, Institute of Modern Physics,
Chinese Academy of Sciences, Lanzhou 730000, China}
\author{U. Lombardo}
\affiliation{INFN-LNS, 44 Via S. Sofia, I-95123 Catania, Italy}

\begin{abstract}
An exact treatment of the operators $Q/e(\omega)$ and the total
momentum is adopted to solve the nuclear matter
Bruecker-Bethe-Goldstone equation with two- and three-body forces.
The single-particle potential, equation of state and nucleon
effective mass are calculated from the exact $G$-matrix. The
results are compared with those obtained under the angle-average
approximation and the angle-average approximation with total
momentum approximation. It is found that the angle-average
procedure, whereas preventing huge calculations of coupled
channels, nevertheless provides a fairly accurate approximation.
On the contrary, the total momentum approximation turns out to be
quite inaccurate compared to its exact counterpart.
\end{abstract}

\maketitle

\section{INTRODUCTION}
One of the central objective in modern nuclear theory is to
understand the saturation properties of nuclear matter, starting
from the $ab$ $inito$ calculation.
Different many-body theories have been developed to settle this
problem, such as the many-body perturbation theory
\cite{mb1,mb2,mb3,mb4,mb5}, the variational method \cite{vm1,vm2},
the Monte Carlo method in its various versions
\cite{mc1,mc2,mc3,mc4,mc5,mc6}, and the diagrammatic expansion
method, in particular, the Brueckner-Bethe-Goldstone (BBG)
hole-line expansion \cite{bhf1,bhf2} and the self-consistent
Green's function approach \cite{scg1,scg2,scg3,scg4}. In the BBG
model, the effect of the nuclear medium is taken into account via
the Pauli operator, which limits the allowed intermediate states
to particle states above the Fermi level, and the self-energy in
the denominator of the two-particle propagator. The accurate
treatment of the Pauli blocking operator and of the denominator of
the two-particle propagator (we call it energy denominator
hereafter) is one of the essential requirements for the numerical
calculation.

The Pauli operator and energy denominator depend, in principle,
not only on the magnitudes of total and relative momenta of the
intermediate two nucleons, but also on their angles. Different
partial waves can couple to each other due to this angular
dependence, leading to cumbersome numerical computations. One can
avoid this difficulty adopting the angle-average (AA)
approximation \cite{aa1,aa2}. Various studies have assessed the
reliability of the AA approximation. Among these, major attention
has been devoted to the angular dependence of the Pauli operator
\cite{q1,q2,q3,q4,q5}, whereas complications arising from the
angular dependence of the energy denominator were handled by the
AA procedure or effective-mass approximation \cite{eaa1}. There
are also attempts to check the angular dependence of the energy
denominator \cite{ee1,ee2}. In Ref. \cite{ee1}, Sartor reports
solutions to the BBG equation with the exact propagator with the
Reid soft-core potential \cite{red}, including the coupling of
different partial waves. It has been concluded that the
corrections due to the angular dependence of the energy
denominator are marginal comparing to the exact treatment of the
Pauli  operator. However, Ref. \cite{ee2} shows the mass operator
and the saturation properties can be affected by taking a monopole
approximation for the propagator with the Argonne $V_{18}$ and
Paris potential. The three-body force (3BF), which is crucial for
reproducing the empirical saturation point of symmetric nuclear
matter \cite{tbf1,tbf2}, was not included in these studies. To
evaluate accurately the AA approximation, the 3BF should be
embodied either.

On the other hand, because of the computational limits, the total
momentum of the intermediate two nucleons was approximated by its
average value in the first calculations Brueckner $et
al$\cite{tma1}. Such an approximation for the total momentum was
widely adopted in previous non relativistic
\cite{tma2,tma3,tma4,tma5,sz1,tma6,tma7} and relativistic
investigations \cite{tmar1,tmar2,tmar3}. However, it is now
possible to refrain from this approximation, and, in fact, a
recent relativistic investigation shows a sizable contribution to
the saturation property \cite{th1} within an exact treatment of
the total momentum. The total momentum approximation in the
nonrelativistic regime has not been investigated yet. In the
present paper we solve the BBG equation exactly, considering the
coupling of different partial waves and abandoning the total
momentum approximations. The reliability of the angular average
and the total momentum approximation will be investigated with the
realistic Argonne $V_{18}$ including also a microscopic 3BF.

In Sec. II we derive the exact partial-wave expansion of BBG
equation. The angle-average and total momentum approximations are
described in Sec. III. The numerical results are presented in Sec.
IV and the summary and outlook are finally given in Sec. V.


\section{THEORETICAL APPROACHES}
The starting point of BBG theory is the Brueckner reaction matrix
$G$, which describes the scattering of two nucleons inside the
nuclear medium. The $G$ matrix satisfies the Bethe-Goldstone
equation,
\begin{eqnarray}
G(\omega)=v +v\frac{Q}{e(\omega)}G(\omega).
\end{eqnarray}
where $Q$, $\omega$, and $e(\omega)$ are the Pauli operator, the
starting energy, and the energy denominator, respectively. The
starting energy is a parameter in the calculation of the various
quantities [e.g., the mass operator $M(k,\omega)$]. In the
previous studies
\cite{q1,q2,q3,q4,q5,bhf1,bhf2,ee1,ee2,tbf2,ebhf}, the Pauli
operator with the energy denominator had been defined as follows:
\begin{eqnarray}
&&\frac{Q}{e(\omega)}|\emph{\textbf{K}}\emph{\textbf{k}}\sigma_{1}\tau_{1}\sigma_{2}\tau_{2}\rangle=
F^{\tau_{1}\tau_{2}}(K,k,\hat{\emph{\textbf{k}}})|\emph{\textbf{K}}\emph{\textbf{k}}\sigma_{1}\tau_{1}\sigma_{2}\tau_{2}\rangle,\nonumber\\
\end{eqnarray}
with the definition
\begin{eqnarray}
&&F^{\tau_{1}\tau_{2}}(K,k,\hat{\emph{\textbf{k}}})\nonumber\\
&=&\frac{[1-n^{\tau_{1}}(|\frac{\emph{\textbf{K}}}{2}+\emph{\textbf{k}}|)][1-n^{\tau_{2}}(|\frac{\emph{\textbf{K}}}{2}-\emph{\textbf{k}}|)]}
{\omega-\frac{\emph{\textbf{K}}^{2}}{4m}-\frac{\emph{\textbf{k}}^{2}}{m}-U^{\tau_{1}}(|\frac{\emph{\textbf{K}}}{2}+\emph{\textbf{k}}|)
-U^{\tau_{2}}(|\frac{\emph{\textbf{K}}}{2}-\emph{\textbf{k}}|)+\imath\eta}.\nonumber\\
\end{eqnarray}

where $\emph{\textbf{K}}$ and $\emph{\textbf{k}}$ (
$k=|\emph{\textbf{k}}|$) are the total and relative momenta of the
scattering nucleons, respectively. The neutron and proton rest
masses are assumed to be equal to the average value $m$ of the
nucleon mass. By $n^{\tau}(k)$ [$\tau=n,p$] we denote the Fermi
distribution function, which at zero temperature is given by the
Heaviside step function
$\theta^{\tau}(\textrm{k}-\textrm{k}^{\tau}_{F})$ with the Fermi
momentum $k^{\tau}_{F}$. The so-called \emph{auxiliary} potential
$U^{\tau}$ is defined as:
\begin{eqnarray}
U^{\tau}(1)=\sum_{\emph{\textbf{p}}'\sigma'\tau'}n^{\tau'}(2)\textrm{Re}\langle
12|G[\varepsilon^{\tau}(1)+\varepsilon^{\tau'}(2)]|12\rangle_{A},
\end{eqnarray}
where $1\equiv(\emph{\textbf{p}},\sigma,\tau)$ denote the
momentum, the spin $z$ component, and the isospin $z$ component of
the particle, respectively. The single-particle (s.p.) energy in
the Brueckner-Hartree-Fock (BHF) approaches reads
\begin{eqnarray}
\varepsilon^{\tau}(\textbf{p})=\frac{\textbf{p}^{2}}{2m}+U^{\tau}(\textbf{p}).
\end{eqnarray}
The \emph{auxiliary} potential $U^{\tau}$ is also called the s.p.
potential in the BHF approaches.

\subsection{Matrix elements of the propagator $Q/e(\omega)$
in the partial-wave expansion}

Usually, the BBG equation is solved in the partial-wave
representation, where the $NN$ interaction can be easily
expressed.
Here $\ell$, $S$, $J$, and $T$ are the orbit angular momentum, the
spin, the total angular momentum, and the isospin of the two
scattering nucleons, respectively. $m_{J}$ and $T_{z}$ are the $z$
components of $J$ and $T$, respectively. This basis can be
expressed as the linear combination of the plane waves
$|\emph{\textbf{K}}\emph{\textbf{k}}\rangle$, i.e,
\begin{eqnarray}
&&|\emph{\textbf{K}}k\ell SJm_{J}TT_{z}\rangle \nonumber\\&=&\int
d\hat{\emph{\textbf{k}}}\sum_{m_{\ell}m_{s}}(\ell
m_{\ell}Sm_{s}|Jm_{J})Y_{\ell
m_{\ell}}(\hat{\emph{\textbf{k}}})|\emph{\textbf{K}}\emph{\textbf{k}}\rangle|Sm_{s}\rangle|TT_{z}\rangle ,\nonumber \\
\end{eqnarray}
here $\hat{\emph{\textbf{k}}}\equiv\ \emph{\textbf{k}}/k$. One
should note that the partial wave basis is not the eigenstate of
the operators $Q/e(\omega)$. [The eigenstate of the Pauli operator
as well as the two-particle propagator should be
$|\emph{\textbf{K}}\emph{\textbf{k}}\sigma_{1}\tau_{1}\sigma_{2}\tau_{2}\rangle$
or equivalently
$|\emph{\textbf{k}}_{1}\sigma_{1}\tau_{1}\emph{\textbf{k}}_{2}\sigma_{2}\tau_{2}\rangle$
with
$\emph{\textbf{k}}_{1,2}=\frac{\emph{\textbf{K}}}{2}\pm\emph{\textbf{k}}$.
With the help of the transformation relation,
\begin{eqnarray}
&&\langle\emph{\textbf{K}}'\emph{\textbf{k}}'\sigma_{1}\tau_{1}\sigma_{2}\tau_{2}|\emph{\textbf{K}}k\ell
SJm_{J}TT_{z}\rangle\nonumber\\
&=&(2\pi)^{3}\delta(\emph{\textbf{K}}-\emph{\textbf{K}}')(2\pi)^{3}\frac{\delta(k-k')}{k^{2}}\nonumber\\
&\times&\sum_{m_{\ell}m_{s}}(\frac{1}{2}\sigma_{1}\frac{1}{2}
\sigma_{2}|Sm_{s})(\frac{1}{2}\tau_{1}\frac{1}{2}
\tau_{2}|TT_{z})\nonumber\\
&\times&(\ell m_{\ell}Sm_{s}|Jm_{J})Y_{\ell
m_{\ell}}(\hat{\emph{\textbf{k}}}'),
\end{eqnarray}
the matrix elements of the propagator $Q/e(\omega)$ is given by
\begin{eqnarray}
&&\langle\emph{\textbf{K}}'k'\ell'
S'J'm_{J}'T'T_{z}'|\frac{Q}{e(\omega)}|\emph{\textbf{K}}k\ell
SJm_{J}TT_{z}\rangle\nonumber\\
&=&\sum_{\emph{\textbf{K}}''\emph{\textbf{k}}''\sigma_{1}\sigma_{2}\tau_{1}\tau_{2}}\langle\emph{\textbf{K}}'k'\ell'
S'J'm_{J}'T'T_{z}'|\emph{\textbf{K}}''\emph{\textbf{k}}''\sigma_{1}\tau_{1}\sigma_{2}\tau_{2}\rangle\nonumber\\
&\times&F^{\tau_{1}\tau_{2}}(K'',k'',\hat{\emph{\textbf{k}}}'')\langle\emph{\textbf{K}}''\emph{\textbf{k}}''\sigma_{1}\tau_{1}\sigma_{2}\tau_{2}|\emph{\textbf{K}}k\ell
SJm_{J}TT_{z}\rangle\nonumber\\
&=&(2\pi)^{3}\delta(\emph{\textbf{K}}-\emph{\textbf{K}}')(2\pi)^{3}\frac{\delta(k-k')}{k^{2}}\delta_{SS'}\nonumber\\
&\times&\sum_{m_{\ell}'m_{\ell}m_{s}\tau_{1}\tau_{2}}(\ell'
m_{\ell}'Sm_{s}|J'm_{J}')(\frac{1}{2}\tau_{1}\frac{1}{2}
\tau_{2}|T'T_{z})\nonumber\\
&\times&(\ell
m_{\ell}Sm_{s}|Jm_{J})(\frac{1}{2}\tau_{1}\frac{1}{2}
\tau_{2}|TT_{z})\nonumber\\
&\times&\int d\hat{\emph{\textbf{k}}}Y^{*}_{\ell'
m_{\ell}'}(\hat{\emph{\textbf{k}}})F^{\tau_{1}\tau_{2}}(K,k,\hat{\emph{\textbf{k}}})Y_{\ell
m_{\ell}}(\hat{\emph{\textbf{k}}}).
\end{eqnarray}
Due to the conservation of the total momentum, the orientation of
the total momentum does not affect the calculation. Its direction
can be chosen as the $z$ axis, therefore, the value of
$F^{\tau_{1}\tau_{2}}$ depends upon the magnitude of the total and
relative momenta of the scattering nucleons, and it is only
affected by the orientation of the relative momentum. Moreover,
the function $F^{\tau_{1}\tau_{2}}$ is axisymmetric along the $z$
axis [the orientation of $\emph{\textbf{K}}$]. In Eq. (8) the
integration over angle $\varphi$ of the spherical coordinates
will, thus, yield a $\delta_{m_{\ell}m_{\ell}'}$ factor, which
allows the $m_{\ell}'$ summation to be performed trivially. The
Clebsch-Gordan coefficients imply that both $m_{J}'$ and $m_{J}$
are equal to $m_{\ell}+m_{s}$, hence, the propagator is diagonal
in $m_{J}$.

Note that $F^{\tau_{1}\tau_{2}}$ [Eq. (3)] also depends upon the
isospin, the summation of Clebsch-Gordan coefficients with
$F^{\tau_{1}\tau_{2}}$ over ${\tau_{1}\tau_{2}}$ could not lead to
the relation $T=T'$ directly. Nevertheless, the summation can be
separated into the symmetric and antisymmetric parts, i.e.,
\begin{eqnarray}
&&\sum_{\tau_{1}\tau_{2}}(\frac{1}{2}\tau_{1}\frac{1}{2}
\tau_{2}|T'T_{z})(\frac{1}{2}\tau_{1}\frac{1}{2}
\tau_{2}|TT_{z})F^{\tau_{1}\tau_{2}}(K,k,\hat{\emph{\textbf{k}}})\nonumber\\
&=&F^{\tau_{1}\tau_{2}}_{S}(K,k,\hat{\emph{\textbf{k}}})|_{\delta_{TT'}}+F^{\tau_{1}\tau_{2}}_{A}(K,k,\hat{\emph{\textbf{k}}})|_{\epsilon_{TT'}},
\end{eqnarray}
where $\epsilon_{TT'}=1-\delta_{TT'}$. And each part corresponds
to a different relationship between $T$ and $T'$ (see the Appendix
for details). Accordingly, the matrix elements of the operators
$Q/e(\omega)$ are calculated as
\begin{eqnarray}
&&\langle\emph{\textbf{K}}'k'\ell'
S'J'm_{J}'T'T_{z}'|\frac{Q}{e(\omega)}|\emph{\textbf{K}}k\ell
SJm_{J}TT_{z}\rangle\nonumber\\
&=&(2\pi)^{3}\delta(\emph{\textbf{K}}-\emph{\textbf{K}}')(2\pi)^{3}\frac{\delta(k-k')}{k^{2}}\delta_{SS'}\delta_{m_{J}m_{J}'}\delta_{T_{z}T_{z}'}\nonumber\\
&\times&\big[\langle k\ell'J'|\mathcal {F}_{S}(K,\omega)|k\ell
J\rangle_{\delta_{TT'}}\nonumber\\
&+&\langle k\ell'J'|\mathcal {F}_{A}(K,\omega)|k\ell
J\rangle_{\epsilon_{TT'}}\big],
\end{eqnarray}
with
\begin{eqnarray}
&&\langle k\ell'J'|\mathcal {F}_{S}(K,\omega)|k\ell J\rangle\nonumber\\
&=&\sum_{m_{\ell}m_{s}}(\ell' m_{\ell}Sm_{s}|J'm_{J})(\ell
m_{\ell}Sm_{s}|Jm_{J})\nonumber\\
&\times&\int d\hat{\emph{\textbf{k}}}Y^{*}_{\ell'
m_{\ell}'}(\hat{\emph{\textbf{k}}})F^{\tau_{1}\tau_{2}}_{S}(K,k,\hat{\emph{\textbf{k}}})Y_{\ell
m_{\ell}}(\hat{\emph{\textbf{k}}})
\end{eqnarray}
and
\begin{eqnarray}
&&\langle k\ell'J'|\mathcal {F}_{A}(K,\omega)|k\ell J\rangle\nonumber\\
&=&\sum_{m_{\ell}m_{s}}(\ell' m_{\ell}Sm_{s}|J'm_{J})(\ell
m_{\ell}Sm_{s}|Jm_{J})\nonumber\\
&\times&\int d\hat{\emph{\textbf{k}}}Y^{*}_{\ell'
m_{\ell}'}(\hat{\emph{\textbf{k}}})F^{\tau_{1}\tau_{2}}_{A}(K,k,\hat{\emph{\textbf{k}}})Y_{\ell
m_{\ell}}(\hat{\emph{\textbf{k}}})
\end{eqnarray}

Once the charge-dependent Argonne $V_{18}$ $NN$ potential is
adopted, the \emph{auxiliary} potential $U^{n}\neq U^{p}$ for
symmetric nuclear matter. Moreover, it is generally true that
$U^{n}\neq U^{p}$ for asymmetric nuclear matter. Consequently, the
definition $F^{pn}(K,k,\hat{\emph{\textbf{k}}})$ is neither an
even function nor an odd function of $\hat{\emph{\textbf{k}}}$.
The AA approximation of $Q/e(\omega)$ in the previous
investigation removes the odd part of
$F^{pn}(K,k,\hat{\emph{\textbf{k}}})$ and the reserved part
ensures the conservation of the parity (see the Appendix). In Ref.
\cite{ee1}, Sartor did not distinguish the neutron and proton
strictly in the derivation, i.e., omitting the antisymmetric part
$F^{pn}_{A}(K,k,\hat{\emph{\textbf{k}}})$. From Eqs. (10) and (12)
the anti-symmetric part $F^{pn}_{A}(K,k,\hat{\emph{\textbf{k}}})$
would result in the mixing of the total isospin $T=0$ and $T=1$
neutron-proton states. Due to the property of
$F^{pn}_{A}(K,k,\hat{\emph{\textbf{k}}})$, this mixing preserves
the generalized Paul principle selection rule $(-1)^{T+S+L}=-1$
(see the Appendix for details). In fact, we have estimated the
effects of the mixing by considering the antisymmetric part
$F^{pn}_{A}(K,k,\hat{\emph{\textbf{k}}})$ in solving the BBG
equation self-consistently, and the results manifest that the
matrix elements of effective $G$ corresponding to the mixing is
less than $0.1~\%$ and the energy due to this mixing is even
smaller.

We stress here the mixing of the total isospin $T=0$ and $T=1$
neutron-proton states, which originates from the definition of
$Q/e(\omega)$ [Eq. (3)], is nonphysical. The definition in Eq. (3)
was first adopted for symmetric nuclear matter with a
charge-independent potential \cite{bhf1,aa1} in which the freedom
of proton and neutron was not considered explicitly. For
asymmetric nuclear matter or the charge-dependent potential
adopted, the definition of $Q/e(\omega)$ should be modified. In
the present paper, we propose a symmetrization of $Q/e(\omega)$,
i.e.,
\begin{eqnarray}
&&F^{\tau_{1}\tau_{2}}(K,k,\hat{\emph{\textbf{k}}})\nonumber\\
&=&\frac{1}{2}\big\{\frac{[1-n^{\tau_{1}}(|\frac{\emph{\textbf{K}}}{2}+\emph{\textbf{k}}|)][1-n^{\tau_{2}}(|\frac{\emph{\textbf{K}}}{2}-\emph{\textbf{k}}|)]}
{\omega-\frac{\emph{\textbf{K}}^{2}}{4m}-\frac{\emph{\textbf{k}}^{2}}{m}-U^{\tau_{1}}(|\frac{\emph{\textbf{K}}}{2}+\emph{\textbf{k}}|)
-U^{\tau_{2}}(|\frac{\emph{\textbf{K}}}{2}-\emph{\textbf{k}}|)+i0}\nonumber\\
&+&\frac{[1-n^{\tau_{1}}(|\frac{\emph{\textbf{K}}}{2}-\emph{\textbf{k}}|)][1-n^{\tau_{2}}(|\frac{\emph{\textbf{K}}}{2}+\emph{\textbf{k}}|)]}
{\omega-\frac{\emph{\textbf{K}}^{2}}{4m}-\frac{\emph{\textbf{k}}^{2}}{m}-U^{\tau_{1}}(|\frac{\emph{\textbf{K}}}{2}-\emph{\textbf{k}}|)
-U^{\tau_{2}}(|\frac{\emph{\textbf{K}}}{2}+\emph{\textbf{k}}|)+i0}\big\},\nonumber\\
\end{eqnarray}
to remove the mixing of total isospin $T=0$ and $T=1$
neutron-proton states. Actually, one might obtain this formula
following Day's derivation \cite{bhf1} by distinguishing the
proton and neutron specifically. Using this definition the
antisymmetric part in Eq. (10) vanishes as well as the mixing.

\subsection{Partial-wave expansion of the Brueckner-Bethe-Goldstone equation}

Using the partial-wave basis, the standard symmetry properties of
the $NN$ interaction are expressed as
\begin{eqnarray}
&&\langle\emph{\textbf{K}}'k'\ell'
S'J'm_{J}'T'T_{z}'|v|\emph{\textbf{K}}k\ell
SJm_{J}TT_{z}\rangle\nonumber\\
&=&(2\pi)^{3}\delta(\emph{\textbf{K}}-\emph{\textbf{K}}')\delta_{SS'}\delta_{JJ'}\delta_{m_{J}m_{J}'}\delta_{TT'}\delta_{T_{z}T_{z}'}\nonumber\\
&\times&\langle k'\ell'JST|v|k\ell JST\rangle.
\end{eqnarray}
From Eqs. (10) and (14) and the closure relation pertaining to the
basis $|\emph{\textbf{K}}k\ell SJm_{J}TT_{z}\rangle$, the BBG
equation can be written in the form
\begin{eqnarray}
&&\langle k'\ell' J'|G(K,\omega)|k\ell
J\rangle\nonumber\\
&=&\delta_{JJ'}\langle k'\ell' J|v|k\ell
J\rangle+\sum_{\ell''\ell'''J'''}\int\frac{k''^{2}dk''}{(2\pi)^{2}}
\langle k'\ell' J'|v|k''\ell'' J'\rangle\nonumber\\
&\times&\langle k''\ell'' J'|\mathcal {F}_{S}(K,\omega)|k''\ell'''
J'''\rangle\langle k''\ell'''
J'''|G(K,\omega)|k\ell J\rangle\nonumber\\
\end{eqnarray}
for fixed $T$, $S$, and $m_{J}$. The total momentum
$\emph{\textbf{K}}$ affects the effective $G$ only by its value
$K$ implicitly. Here the invariants, i.e., $\emph{\textbf{K}}$,
$T$, $S$, $m_{J}$ and $T_{z}$, have not been written out
explicitly in the expression. The same as in Refs. \cite{q4,ee1}
for fixed $ST$ channels, there are coupling between various total
momenta $J$'s.

The \emph{auxiliary} potential $U^{\tau}$ can be expressed as
$U^{\tau}\equiv
U^{\tau_{1}}_{\sigma_{1}}=\sum_{\tau_{2}}U^{\tau_{1}\tau_{2}}_{\sigma_{1}}$
with
\begin{eqnarray}
&&U^{\tau_{1}\tau_{2}}_{\sigma_{1}}(k_{1})\nonumber\\
&=&2\int\frac{d\emph{\textbf{k}}_{2}}{(2\pi)^{3}}
\sum_{TSm_{J}}\sum_{\ell'J'\ell J}\textrm{Re}\langle k\ell'
J'|G(K,E_{2})|k\ell J\rangle|_{TSm_{J}T_{z}} \nonumber\\
&\times&\imath^{\ell'-\ell}\sum_{\sigma_{2}}(\frac{1}{2}
\sigma_{1}\frac{1}{2} \sigma_{2}|Sm_{s})^{2}(\frac{1}{2}
\tau_{1}\frac{1}{2}\tau_{2}|TT_{z})^{2}\nonumber\\
&\times&(\ell' m_{\ell}Sm_{s}|J'm_{J})Y_{\ell'
m_{\ell}}^{*}(\hat{\emph{\textbf{k}}})(\ell
m_{\ell}Sm_{s}|Jm_{J})Y_{\ell
m_{\ell}}(\hat{\emph{\textbf{k}}}),\nonumber\\
\end{eqnarray}
where
$E_{2}=\varepsilon^{\tau_{1}}(\emph{\textbf{k}}_{1})+\varepsilon^{\tau_{2}}(\emph{\textbf{k}}_{2})$.
For fixed $\sigma_{1}$, $m_{s}=\sigma_{1}+\sigma_{2}$, and
$m_{\ell}=m_{J}-m_{s}$. In the integral of
$d\Omega_{\emph{\textbf{k}}_{2}}$, one should note that
$\cos\theta_{\emph{\textbf{k}}}=\frac{k_{1}-k_{2}\cos\theta_{\emph{\textbf{k}}_{2}}}{2k}$
and $\varphi_{\emph{\textbf{k}}}=\varphi_{\emph{\textbf{k}}_{2}}$
with
$\hat{\emph{\textbf{k}}}_{2}=(\theta_{\emph{\textbf{k}}_{2}},\varphi_{\emph{\textbf{k}}_{2}})$
and
$\hat{\emph{\textbf{k}}}=(\theta_{\emph{\textbf{k}}},\varphi_{\emph{\textbf{k}}})$.
In the exact calculations, the BBG equation (15), the
\emph{auxiliary} potential potential (16), and the s.p. energy (5)
are solved self-consistently by taking the off-diagonal matrix
elements of $\langle\ell' J'|G|\ell J\rangle$.

\subsection{The angle-average approximation and the total momentum approximation}
In the AA procedure, $F^{\tau_{1}\tau_{2}}$ is replaced by its
averaged value, i.e.,
\begin{eqnarray}
F^{\tau_{1}\tau_{2}}(K,k,\hat{\emph{\textbf{k}}}) \longrightarrow
\overline{F^{\tau_{1}\tau_{2}}}(K,k)\equiv\int
\frac{d\Omega_{\emph{\textbf{k}}}}{4\pi}F^{\tau_{1}\tau_{2}}(K,k,\hat{\emph{\textbf{k}}}).\nonumber\\
\end{eqnarray}
Thus, the integral in Eq. (16) yields $\delta_{\ell\ell'}$, and
the summation over $m_{\ell}$ and $m_{s}$ gives $\delta_{JJ'}$.
Thus, the matrix elements of the operators $Q/e(\omega)$ are
written as
\begin{eqnarray}
&&\langle\emph{\textbf{K}}'k'\ell'
S'J'm_{J}'T'T_{z}'|\frac{Q}{e(\omega)}|\emph{\textbf{K}}k\ell
SJm_{J}TT_{z}\rangle\nonumber\\
&=&(2\pi)^{3}\delta(\emph{\textbf{K}}-\emph{\textbf{K}}')(2\pi)^{3}
\frac{\delta(k-k')}{k^{2}}\nonumber\\
&\times&\overline{F^{\tau_{1}\tau_{2}}}(K,k)\delta_{SS'}\delta_{JJ'}\delta_{\ell\ell'}
\delta_{m_{J}m_{J}'}\delta_{T_{z}T_{z}'}.
\end{eqnarray}
The BBG equation can be simplified as
\begin{eqnarray}
&&\langle k'\ell' J|G(K,\omega)|k\ell
J\rangle\nonumber\\
&=&\langle k'\ell' J|v|k\ell
J\rangle+\sum_{\ell''}\int\frac{k''^{2}dk''}{(2\pi)^{2}}
\langle k'\ell' J|v|k''\ell'' J\rangle\nonumber\\
&\times&\overline{F^{\tau_{1}\tau_{2}}}(K,k)\langle k''\ell''
J|G(K,\omega)|k\ell J\rangle.
\end{eqnarray}
Here the coupling between different partial waves is eliminated,
only a coupling between different orbital angular momenta $\ell$'s
due to the tensor force. The \emph{auxiliary} potential is
simplified as
\begin{eqnarray}
&&U^{\tau_{1}\tau_{2}}_{\sigma_{1}}(k_{1})\nonumber\\
&=&\sum_{JS\ell}\frac{(2J+1)(1+\delta_{\tau_{1}\tau_{2}})}{4}
\int\frac{k_{2}^{2}dk_{2}}{(2\pi)^{3}}
 \nonumber\\
&\times&\int\sin\theta_{\emph{\textbf{k}}_{2}}d\theta_{\emph{\textbf{k}}_{2}}
\textrm{Re}\langle k\ell J|G(K,E_{2})|k\ell J\rangle.
\end{eqnarray}
Equations (5), (19), and (20) are essentially solved
self-consistently in the AA approximation.

The calculations of the \emph{auxiliary} potential, i.e., Eq.
(20), need the full information of $G$ at arbitrary values of $K$
and $\omega$. One actually solves the BBG equation on a
$N_{K}\times N_{\omega}$ grid, where $N_{K} (N_{\omega})$ is the
number of the $K (\omega)$ points. Several decades ago, such
calculations were greatly challenging. Consequently, the total
momentum approximation has been adopted \cite{tma1}, which is
defined as
\begin{eqnarray}
&&\langle K_{\tau\tau'}^{2}\rangle(k)\nonumber\\&=&\frac{\int
d\emph{\textbf{k}}_{1}\int
d\emph{\textbf{k}}_{2}n^{\tau}(k_{1})n^{\tau'}(k_{1})
\emph{\textbf{K}}^{2}\delta(k-|\emph{\textbf{k}}_{1}-\emph{\textbf{k}}_{2}|/2)}{\int
d\emph{\textbf{k}}_{1}\int
d\emph{\textbf{k}}_{2}n^{\tau}(k_{1})n^{\tau'}(k_{1})
\delta(k-|\emph{\textbf{k}}_{1}-\emph{\textbf{k}}_{2}|/2)}. \ \
\nonumber\\
\end{eqnarray}
In the present paper, the total momentum approximation (TMA)
refers to adopting this average value of total momentum.

In the 3BF model adopted here, the most important mesons, i.e.,
$\pi$, $\rho$, $\sigma$, and $\omega$ have been considered. Using
the one-boson-exchange potential model, all the parameters of the
3BF model, i.e., the coupling constants and the form factors, are
self-consistently determined to reproduce the Argonne $V_{18}$
potential, and their values can be found in Ref. \cite{tbf2}.
After a suitable integration over the degrees of freedom of the
third nucleon, the 3BF can be reduced to an equivalent effective
two-body force according to the standard scheme as described in
Ref. \cite{tbf1}. The equivalent two-body force
$V_{3}^{\text{eff}}$ in $r$ space reads
\begin{eqnarray}
&&\langle\emph{\textbf{r}}_{1}',\emph{\textbf{r}}_{2}'|V_{3}^{eff}|\emph{\textbf{r}}_{1},\emph{\textbf{r}}_{2}\rangle\nonumber\\
&=&\frac{1}{4}Tr\sum_{n}\int
d\emph{\textbf{r}}_{3}'\emph{\textbf{r}}_{3}\phi_{n}^{*}(\emph{\textbf{r}}_{3}')
[1-\eta(r_{13}')] [1-\eta(r_{23}')]\nonumber\\
&\times&W_{3}(\emph{\textbf{r}}_{1}',\emph{\textbf{r}}_{2}',\emph{\textbf{r}}_{3}'|\emph{\textbf{r}}_{1},\emph{\textbf{r}}_{2},\emph{\textbf{r}}_{3})
\phi_{n}(\emph{\textbf{r}}_{3})[1-\eta(r_{13})][1-\eta(r_{23})],\nonumber\\
\end{eqnarray}
where $\phi_{n}$ is the wave function of the single nucleon in
free space and the trace is taken with respect to the spin and
isospin of the third nucleon. $W_{3}$ represents the 3BF as
described in Ref. \cite{tbf1}. Note that the averaging procedure
of the third nucleon , which avoids the difficult problem to solve
the Faddeev equation involving the 3BF, neglects certain many-body
contributions \cite{3bf1,3bf2}. The defect function $\eta(r)$ is
directly related to the $G$ matrix and should be calculated
self-consistently with the BBG equation. Also, the defect function
implicitly depends on the value of total momentum. On the
contrary, in the TMA, the total momenta are approximated by the
average values for both $G$ matrix and the defect function. We
stress here, in our exact numerical treatment, we do not use the
total momentum approximation in the determination of the defect
function as well.


\section{RESULTS AND DISCUSSION}
\begin{figure}[htb!]
\centering
\includegraphics[width=1.0\linewidth]{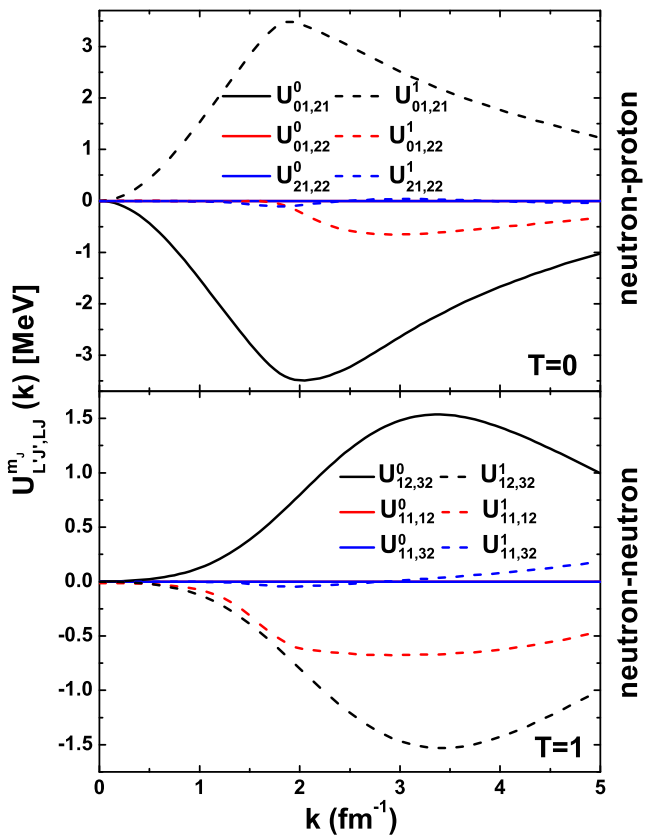} \caption{
Nuclear potential as calculated from the off-diagonal matrix
elements of $G$ matrix with the angular dependence of the
propagator $Q/e(\omega)$. The upper (lower) panel corresponds to
the neutron-proton (neutron-neutron) in the isospin singlet
(triplet) state.} \label{off}
\end{figure}
Compared to the AA, two primary changes stem from the angular
dependence of the operator $Q/e(\omega)$. First, the off-diagonal
matrix elements of $G$ matrix in $J$, i.e., $\langle
J'|G|J\rangle\neq0$ [here the other variables are omitted]
represent the coupling of different channels with  fixed spin and
isospin. These off-diagonal $G$ matrix elements result in a non
vanishing contribution to the s.p. potentials. In the upper panel
of Fig. 1 we show $U_{\ell'J',\ell J}^{m_{J}}(k)$ [which is
obtained calculating Eq. (16) without the  $TS\ell Jm_{j}$
summations] for the isospin-singlet neutron-proton scattering for
symmetric nuclear matter at the empirical saturation density
$\rho_{0}=0.17\text{fm}^{-3}$. In this figure, three kinds
couplings, i.e., ($\ell'=0, J'=1;\ell=2,J=1$), ($\ell'=0,
J'=1;\ell=2,J=2$) and ($\ell'=2, J'=1;\ell=2,J=2$), with
$m_{J}=0,1$ are displayed. The sizable coupling ($\ell'=0,
J'=1;\ell=2,J=1$) actually corresponds to the tensor force in the
$^{3}SD_{1}$ channel which contains three $m_{j}$ components. For
spin-up, i.e, $\sigma_{1}=\frac{1}{2}$, the Clebsch-Gordan
coefficients and the properties of the spherical harmonics imply
that the $m_{j}=-1$ components $U_{01,21}^{-1}(k)\equiv0$. In
addition, the Clebsch-Gordan coefficients and the properties of
the spherical harmonics in Eq. (16) render the opposite values for
the two components $U_{01,21}^{0}(k)$ and $U_{01,21}^{1}(k)$.
However, they do not cancel each other completely since the $G$
matrix is nondegenerate for different $m_{j}$'s, that is,
essentially originated from the angular dependence of the
operators. This is the second major change when the AA
approximation is not adopted. Due to the symmetries of $G$ matrix
$\langle\ell'J'|G|\ell
J\rangle_{-m_{J}}=(-)^{\ell'-J'+\ell-J}\langle\ell'J'|G|\ell
J\rangle_{m_{J}}$, $U_{01,22}^{0}(k)$ and $U_{21,22}^{0}(k)$ are
both equal to 0. The significant potential $U_{01,22}^{1}(k)$
corresponding to the coupling between $^{3}S_{1}$ and $^{3}D_{2}$
partial waves indicates that certain couplings, unexpected in the
AA approximation, might contribute to some extent  to the s.p.
potential. For neutron-neutron, except the specific values, they
are similar to the neutron-proton case shown in the lower panel of
Fig.1.

\begin{figure}[htb!]
\centering
\includegraphics[width=1.0\linewidth]{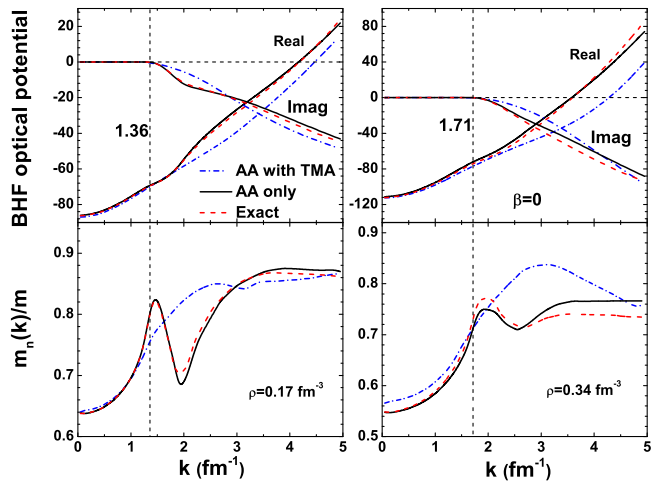} \caption{
BHF optical potential and  neutron effective-mass $m_{n}^{*}/m$ as
a function of the momentum $k$ for symmetric nuclear matter at two
densities with the exact calculation, the AA approximation, and
the AA approximation with TMA, respectively.} \label{uk}
\end{figure}
To investigate the changes in the s.p. potential due to the
angular dependence of the propagator, we report in the upper
panels of Fig. 2 the BHF optical potential of symmetric nuclear
matter in two cases with (solid line) and without (dashed line)
the AA approximation for two typical densities $\rho_{0}$ and
$2\rho_{0}$ , respectively. [Here,the BHF optical potential
represent the on-shell value of Eq. (4).] Moreover, the results
with the AA approximation including TMA, widely used in the
previous works \cite{tbf2,tma1,tma2,tma3,tma4,tma5}, are reported
as well. In calculating the BHF optical potential, one can
actually employ Eq. (16) by considering the imaginary part of $G$
matrix as well as the real part, and $\textrm{Im}G$ corresponds to
the imaginary part of the BHF optical potential. In the lower
panels, the neutron effective mass vs. momentum , which are
related to the derivative of the s.p. potential by
$\frac{m_{\tau}^{*}}{m}(k)=\frac{k}{m}
[\frac{d\varepsilon^{\tau}(k)}{dk}]^{-1}$, is also plotted. At the
empirical saturation density $\rho_{0}$, compared to the exact
scheme, the deviation resulting from the AA approximation is tiny
for both potential and effective mass. This deviation grows up
when increasing density, since the Pauli blocking effects become
stronger and stronger. The procedure of handling with the
$Q/e(\omega)$ becomes more important due to enhancing the Pauli
blocking effect. When the TMA is also adopted, the potential
deviation from the exact results  is quite evident, especially for
large momenta. Fortunately, the deviation is still tolerable below
the Fermi momentum. The same as with the AA approximation, the
deviation becomes more distinct at larger densities, due to the
enhancement of the Pauli blocking. The effective mass in the AA
approximation and TMA manifest a diversity similar to what happens
with the potential. However, the difference looks more remarkable
for the effective mass.

\begin{figure}
\includegraphics[width=1.0\linewidth]{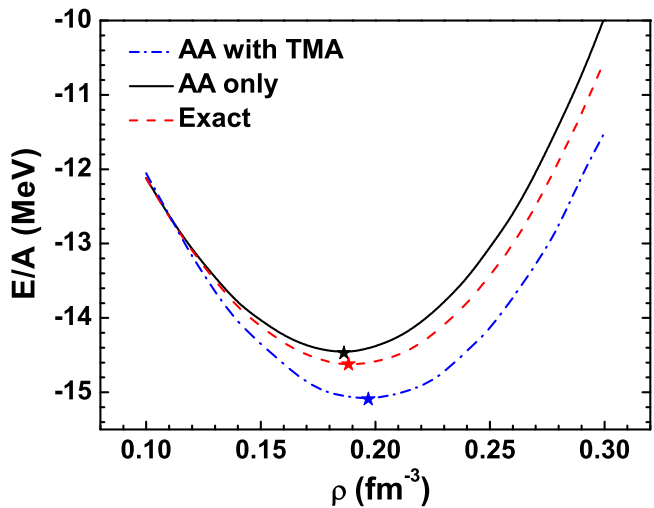} \caption{
Equation of state (EOS) of symmetric nuclear matter with the exact
calculation, the AA approximation, and the AA approximation with
TMA, respectively. The star marks the saturation point in the
three calculations.} \label{upd}
\end{figure}

In Fig. 3. we show the effects of the AA and TMA approximations on
the EOS for symmetric nuclear matter.  The saturation point
($\rho_{s}=0.197 \text{fm}^{-3}, E/A=-15.09 \text{MeV}$) in the AA
approximation with TMA coincides with that of Ref. \cite{tbf2}.
The exact treatment of the total momentum substantially improves
the saturation point. The saturation point within the AA
approximation is about ($\rho_{s}=0.186 \text{fm}^{-3}, E/A=-14.47
\text{MeV}$) approaching that of the exact calculation,
$\rho_{s}=0.188 \text{fm}^{-3}, E/A=-14.62 \text{MeV}$. As
expected,  AA approximation only leads to a small deviation of the
EOS from the exact one. Although the discrepancy between the EOS
of the exact calculation and the AA approximation plus TMA remains
substantial.

\begin{figure}
\includegraphics[width=1.0\linewidth]{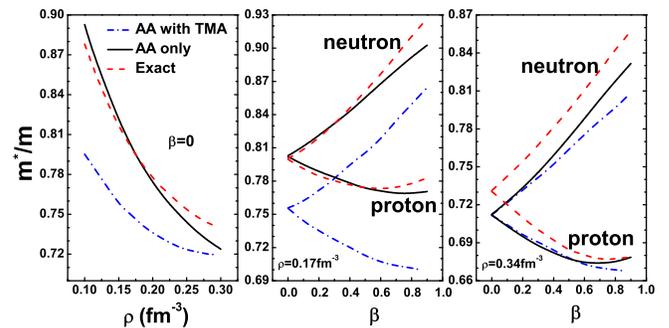} \caption{(Color online).
Nucleon effective mass as a function of the density $\rho$ and the
isospin asymmetry $\beta$ within the exact calculation, the AA
approximation, and the AA approximation with TMA, respectively.}
\label{ems}
\end{figure}
The nucleon effective mass is an important microscopic input to
many studies of nuclear phenomena, such as the dynamics of heavy
ion collisions at intermediate and high energies \cite{em1}, the
damping of nuclear excitations and giant resonances \cite{em2},
and the thermal properties of neutron stars
\cite{em3,em4,em5,em6}. Very important is the value of the
effective mass at the Fermi surface, i.e., $m^{*}(k_{F})$
[hereafter the effective mass is referred to as $m^{*}(k_{F})$].
The density and isospin-asymmetry dependence of the effective mass
is reported in Fig. 4 where the asymmetry parameter is defined as
$\beta=(\rho_{n}-\rho_{p})/(\rho_{n}+\rho_{p})$ with the neutron
(proton) density $\rho_{n}$ ($\rho_{p}$). The results in the AA
approximation with TMA is consistent with that of Ref. \cite{mr1}.
Above all, the effective-mass calculation with AA agrees well with
the exact one at low density, but the discrepancy between the two
grows with increasing density, being the enhancement attributed to
the Pauli blocking effect. Similar to the potential, the
difference between the exact and the AA approximation plus TMA
seems noteworthy, although the difference tends to diminish at
increasing densities. However, one should note that the proton
effective mass decreases monotonically with the isospin asymmetry
when the TMA is adopted. The proton effective mass in the exact
calculation and AA  exhibits a parabolic dependence on the isospin
asymmetry.



\section{SUMMARY}
The equation of state of nuclear matter plays an important role in
modern nuclear physics and astrophysics. The
Brueckner-Bethe-Goldstone theory allows an \emph{ab initio}
calculation from microscopic potential. In the present
investigation, an exact numerical treatment of the BG equation is
performed to assess the reliability of the angular-average
procedure of the operators $Q/e(\omega)$ as well as the total
momentum approximation. In the exact treatment of the Pauli
propagator, the previous definition of $Q/e(\omega)$ Eq.(3), which
was adopted for symmetric nuclear matter, is proved to be
inappropriate for asymmetric nuclear matter since it may lead to a
mixing of the total isospin  T=1 and T=0 neutron-proton states in
$G$-matrix.
Instead, we propose to symmetrize the Pauli propagator, i.e., Eq.
(13), in our calculation to extend the definition of $Q/e$ for
asymmetric matter, consistently with the symmetry
$G(\beta)=G(-\beta)$. Accordingly we compare the s.p. potential,
nucleon effective mass, and EOS in the three calculations with a
microscopic 3BF, i.e., the exact calculation, the AA
approximation, and the AA approximation plus TMA.

The calculations show that the exact treatment of the Pauli
propagator results in inappreciable changes in the potential,
effective mass, and binding energy at low densities. These changes
grow slightly with density due to the enhancement of Pauli
blocking. However, the AA procedure provides a fairly accurate
simplifying approximation even when 3BF is applied. Instead, the
total momentum of the intermediate two nucleons leads to a
considerable contribution in predicting the saturation properties,
the s.p. potential, and nucleon effective mass. And its
replacement TMA is insufficient for accurate investigations of
nuclear matter.

\section*{Acknowledgments}
{We are grateful to Dr. P. Wang for the useful discussion. The
work was supported by the National Natural Science Foundation of
China (Grants No. 11975282, No. 11775276, No. 11435014, and No.
11505241), the Strategic Priority Research Program of the Chinese
Academy of Sciences, Grant No. XDB34000000, the Youth Innovation
Promotion Association of the Chinese Academy of Sciences.}
\appendix

\section{Appendix}
Let us consider, in fact, neutron-proton or proton-neutron matrix
elements with $T_{z}=0$ in Eq. (8). When $T=0$, the summation over
${\tau_{1}\tau_{2}}$ reads
\begin{eqnarray*}
&&\sum_{\tau_{1}\neq\tau_{2}}(\frac{1}{2}\tau_{1}\frac{1}{2}
\tau_{2}|T'T_{z})(\frac{1}{2}\tau_{1}\frac{1}{2}
\tau_{2}|TT_{z})F^{\tau_{1}\tau_{2}}(K,k,\hat{\emph{\textbf{k}}})\big|_{T'=0}\nonumber\\
&=&(\frac{1}{2}\frac{1}{2}\frac{1}{2}
\frac{-1}{2}|00)(\frac{1}{2}\frac{1}{2}\frac{1}{2}
\frac{-1}{2}|00)F^{pn}(K,k,\hat{\emph{\textbf{k}}})|_{T'=0}\nonumber\\
&+&(\frac{1}{2}\frac{-1}{2}\frac{1}{2}
\frac{1}{2}|00)(\frac{1}{2}\frac{-1}{2}\frac{1}{2}
\frac{1}{2}|00)F^{np}(K,k,\hat{\emph{\textbf{k}}})|_{T'=0}\nonumber\\
&=&\frac{1}{2}[F^{pn}(K,k,\hat{\emph{\textbf{k}}})+F^{np}(K,k,\hat{\emph{\textbf{k}}})]|_{T'=0}\nonumber\\
&=&F^{pn}_{S}(K,k,\hat{\emph{\textbf{k}}})|_{T'=0},
\end{eqnarray*}
\begin{eqnarray}
&&\sum_{\tau_{1}\neq\tau_{2}}(\frac{1}{2}\tau_{1}\frac{1}{2}
\tau_{2}|T'T_{z})(\frac{1}{2}\tau_{1}\frac{1}{2}
\tau_{2}|TT_{z})F^{\tau_{1}\tau_{2}}(K,k,\hat{\emph{\textbf{k}}})\big|_{T'=1}\nonumber\\
&=&(\frac{1}{2}\frac{1}{2}\frac{1}{2}
\frac{-1}{2}|10)(\frac{1}{2}\frac{1}{2}\frac{1}{2}
\frac{-1}{2}|00)F^{pn}(K,k,\hat{\emph{\textbf{k}}})|_{T'=1}\nonumber\\
&+&(\frac{1}{2}\frac{-1}{2}\frac{1}{2}
\frac{1}{2}|10)(\frac{1}{2}\frac{-1}{2}\frac{1}{2}
\frac{1}{2}|00)F^{np}(K,k,\hat{\emph{\textbf{k}}})|_{T'=1}\nonumber\\
&=&\frac{1}{2}[F^{pn}(K,k,\hat{\emph{\textbf{k}}})-F^{np}(K,k,\hat{\emph{\textbf{k}}})]|_{T'=1}\nonumber\\
&=&F^{pn}_{A}(K,k,\hat{\emph{\textbf{k}}})|_{T'=1},
\end{eqnarray}
with the symmetric and antisymmetric parts of
$F^{\tau_{1}\tau_{2}}$,
\begin{eqnarray}
 F^{pn}_{S}(K,k,\hat{\emph{\textbf{k}}})
 =\frac{1}{2}[F^{pn}(K,k,\hat{\emph{\textbf{k}}})+F^{pn}(K,k,-\hat{\emph{\textbf{k}}})]\nonumber\\
 F^{pn}_{A}(K,k,\hat{\emph{\textbf{k}}})
 =\frac{1}{2}[F^{pn}(K,k,\hat{\emph{\textbf{k}}})-F^{pn}(K,k,-\hat{\emph{\textbf{k}}})].
\end{eqnarray}
In deriving Eq. (A1), the properties of $F^{\tau_{1}\tau_{2}}$,
\begin{eqnarray}
F^{\tau_{1}\tau_{2}}(K,k,\hat{\emph{\textbf{k}}})=F^{\tau_{2}\tau_{1}}(K,k,-\hat{\emph{\textbf{k}}})
\end{eqnarray}
should be adopted. For the case of $T=1$ and $T_{z}=0$ with the
same procedure we get
\begin{eqnarray*}
&&\sum_{\tau_{1}\neq\tau_{2}}(\frac{1}{2}\tau_{1}\frac{1}{2}
\tau_{2}|T'T_{z})(\frac{1}{2}\tau_{1}\frac{1}{2}
\tau_{2}|TT_{z})F^{\tau_{1}\tau_{2}}(K,k,\hat{\emph{\textbf{k}}})\big|_{T'=1}\nonumber\\
&=&F^{pn}_{S}(K,k,\hat{\emph{\textbf{k}}})|_{T'=1}
\end{eqnarray*}
\begin{eqnarray}
&&\sum_{\tau_{1}\neq\tau_{2}}(\frac{1}{2}\tau_{1}\frac{1}{2}
\tau_{2}|T'T_{z})(\frac{1}{2}\tau_{1}\frac{1}{2}
\tau_{2}|TT_{z})F^{\tau_{1}\tau_{2}}(K,k,\hat{\emph{\textbf{k}}})\big|_{T'=0}\nonumber\\
&=&F^{pn}_{A}(K,k,\hat{\emph{\textbf{k}}})|_{T'=0}.
\end{eqnarray}
Equations (A1) and (A4) indicate that
\begin{eqnarray}
&&\sum_{\tau_{1}\tau_{2}}(\frac{1}{2}\tau_{1}\frac{1}{2}
\tau_{2}|T'T_{z})(\frac{1}{2}\tau_{1}\frac{1}{2}
\tau_{2}|TT_{z})F^{\tau_{1}\tau_{2}}(K,k,\hat{\emph{\textbf{k}}})\nonumber\\
&=&F^{\tau_{1}\tau_{2}}_{S}(K,k,\hat{\emph{\textbf{k}}})|_{\delta_{TT'}}+F^{\tau_{1}\tau_{2}}_{A}(K,k,\hat{\emph{\textbf{k}}})|_{\epsilon_{TT'}}
\end{eqnarray}
for the case of $T_{z}=0$, where $\epsilon_{TT'}=1-\delta_{TT'}$.
Once the proton-proton (neutron-neutron) matrix element with
$T_{z}=1 (-1)$ is calculated, Eq. (A5) is true as well with
$F^{\tau_{1}\tau_{1}}_{A}\equiv0$.

The property of
$F^{\tau_{1}\tau_{2}}_{S}(K,k,\hat{\emph{\textbf{k}}})=F^{\tau_{1}\tau_{2}}_{S}(K,k,-\hat{\emph{\textbf{k}}})$
together with the well-known properties of the spherical
harmonics, implies the identities,
\begin{eqnarray}
&&\int d\hat{\emph{\textbf{k}}}Y^{*}_{\ell'
m_{\ell}'}(\hat{\emph{\textbf{k}}})F^{\tau_{1}\tau_{2}}_{S}(K,k,\hat{\emph{\textbf{k}}})Y_{\ell
m_{\ell}}(\hat{\emph{\textbf{k}}})\nonumber\\
&=&\int d\hat{\emph{\textbf{k}}}Y^{*}_{\ell'
m_{\ell}'}(-\hat{\emph{\textbf{k}}})F^{\tau_{1}\tau_{2}}_{S}(K,k,-\hat{\emph{\textbf{k}}})Y_{\ell
m_{\ell}}(-\hat{\emph{\textbf{k}}})\nonumber\\
&=&(-1)^{\ell'+\ell}\int d\hat{\emph{\textbf{k}}}Y^{*}_{\ell'
m_{\ell}'}(\hat{\emph{\textbf{k}}})F^{\tau_{1}\tau_{2}}_{S}(K,k,\hat{\emph{\textbf{k}}})Y_{\ell
m_{\ell}}(\hat{\emph{\textbf{k}}}),\nonumber\\
\end{eqnarray}
which conserves the parity. On the contrary, the term,
\begin{eqnarray}
&&\int d\hat{\emph{\textbf{k}}}Y^{*}_{\ell'
m_{\ell}'}(\hat{\emph{\textbf{k}}})F^{\tau_{1}\tau_{2}}_{A}(K,k,\hat{\emph{\textbf{k}}})Y_{\ell
m_{\ell}}(\hat{\emph{\textbf{k}}})\nonumber\\
&=&\int d\hat{\emph{\textbf{k}}}Y^{*}_{\ell'
m_{\ell}'}(-\hat{\emph{\textbf{k}}})F^{\tau_{1}\tau_{2}}_{A}(K,k,-\hat{\emph{\textbf{k}}})Y_{\ell
m_{\ell}}(-\hat{\emph{\textbf{k}}})\nonumber\\
&=&-(-1)^{\ell'+\ell}\int d\hat{\emph{\textbf{k}}}Y^{*}_{\ell'
m_{\ell}'}(\hat{\emph{\textbf{k}}})F^{\tau_{1}\tau_{2}}_{A}(K,k,\hat{\emph{\textbf{k}}})Y_{\ell
m_{\ell}}(\hat{\emph{\textbf{k}}}),\nonumber\\
\end{eqnarray}
violates the parity conservation. Fortunately, with the help of
$\epsilon_{TT'}$ and $\delta_{SS'}$ one can demonstrate that this
term maintains the generalized Pauli principle selection rule
$(-1)^{T+S+\ell}=-1$.


\begin{thebibliography}{90}
\bibitem{mb1}
K. Hebeler, S. K. Bogner, R. J. Furnstahl, A. Nogga, and A.
Schwenk, Phys. Rev. C {\bf 83}, 031301(R) (2011).
\bibitem{mb2}
I. Tews, T. Kr$\ddot{u}$ger, K. Hebeler, and A. Schwenk, Phys.
Rev. Lett. {\bf 110}, 032504 (2013).
\bibitem{mb3}
J.W. Holt, N. Kaiser, and W. Weise, Prog. Part. Nucl. Phys. {\bf
73}, 35 (2013).
\bibitem{mb4}
C. Wellenhofer, J.W. Holt, and N. Kaiser, Phys. Rev. C {\bf 92},
015801 (2015).
\bibitem{mb5}
C. Drischler, K. Hebeler, and A. Schwenk, Phys. Rev. Lett. {\bf
122}, 042501 (2019).
\bibitem{vm1}
V. R. Pandharipande and R. B. Wiringa, Rev. Mod. Phys. {\bf 51},
821 (1979).
\bibitem{vm2}
F. Arias de Saavedra, C. Bisconti, G. Co', and A. Fabrocini, Phys.
Rep. {\bf 450}, 1 (2007).
\bibitem{mc1}
B. S. Pudliner, V. R. Pandharipande, J. Carlson, and R. B.
Wiringa, Phys. Rev. Lett. {\bf 74}, 4396 (1995).
\bibitem{mc2}
K. E. Schmidt and S. Fantoni, Phys. Lett. B {\bf 446}, 99 (1999).
\bibitem{mc3}
S. C. Pieper and R. B. Wiringa, Annu. Rev. Nucl. Part. Sci. {\bf
51}, 53 (2001).
\bibitem{mc4}
J. Carlson, J. Morales Jr., V. R. Pandharipande, and D. G.
Ravenhall, Phys. Rev. C {\bf 68}, 025802 (2003).
\bibitem{mc5}
A. Gezerlis and J. Carlson, Phys. Rev. C {\bf 81}, 025803 (2010).
\bibitem{mc6}
G. Wlazlowski and P. Magierski, Phys. Rev. C {\bf 83}, 012801
(2011).
\bibitem{bhf1}
B.D. Day, Rev. Mod. Phys. {\bf 39}, 719 (1967).
\bibitem{bhf2}
M. Baldo, I. Bombaci, G. Giansiracusa, U. Lombardo, C. Mahaux, and
R. Sartor, Phys. Rev. C {\bf 41} 1748 (1990); Nucl. Phys. A {\bf
545} 741 (1992).
\bibitem{scg1}
A. Ramos, A. Polls, and W. Dickoff, Nucl. Phys. A {\bf 551}, 45
(1993).
\bibitem{scg2}
T. Alm et al., Nucl. Phys. A {\bf 551}, 45 (1993).
\bibitem{scg3}
A. Rios, A. Polls, A. Ramos, and H. Muther, Phys. Rev. C {\bf 74},
054317 (2006).
\bibitem{scg4}
A. Rios, A. Polls, and I.Vidana, Phys. Rev. C {\bf 79}, 025802
(2009).
\bibitem{aa1}
K.A. Brueckner, J.L. Gammel, Phys. Rev. {\bf 109}, 1023 (1958).
\bibitem{aa2}
M. I. Haftel, F. Tabakin, Nucl. Phys. A {\bf 158}, 1 (1970).
\bibitem{q1}
T. Cheon and E. F. Redish, Phys. Rev. C {\bf 39}, 331 (1989).
\bibitem{q2}
E. Schiller, H. M¡§uther, and P. Czerski, Phys. Rev. C {\bf 59},
2934 (1999); {\bf 60}, 059901(E) (1999).
\bibitem{q3}
K. Suzuki, R. Okamoto, M. Kohno, and S. Nagata, Nucl. Phys. A {\bf
665}, 92 (2000).
\bibitem{q4}
F. Sammarruca, X. Meng, and E. J. Stephenson, Phys. Rev. C {\bf
62}, 014614 (2000).
\bibitem{q5}
E. J. Stephenson, R. C. Johnson, and F. Sammarruca, Phys. Rev. C
{\bf 71}, 014612 (2005).
\bibitem{eaa1}
T. Frick, Kh. Gad, H. M¡§uther, and P. Czerski, Phys. Rev. C 65,
034321 (2002).
\bibitem{ee1}
R. Sartor, Phys. Rev. C {\bf 54}, 809 (1996).
\bibitem{ee2}
H. F. Arellano, and J.-P. Delaroche, Phys. Rev. C {\bf 83}, 044306
(2011).
\bibitem{red}
R. V. Reid, Ann. Phys. (N.Y.) {\bf 50}, 411 (1968).
\bibitem{tbf1}
P. Grang\'e, A. Lejeune, M. Martzolff, and J.-F. Mathiot, Phys.
Rev. C {\bf 40}, 1040 (1989).
\bibitem{tbf2}
W. Zuo, A. Lejeune, U. Lombardo, and J.-F. Mathiot, Nucl. Phys. A
{\bf 706}, 418 (2002); Eur. Phys. J. A {\bf 14}, 469 (2002).
\bibitem{tma1}
K. A. Brueckner, S. A. Coon, and J. Dabrowski, Phys. Rev. {\bf
168}, 1184 (1968).
\bibitem{tma2}
I. Bombaci and U. Lombardo, Phys. Rev. C {\bf 44}, 1892 (1991).
\bibitem{tma3}
W. Zuo, I. Bombaci, and U. Lombardo, Phys. Rev. C {\bf 60}, 024605
(1999).
\bibitem{tma4}
I. Bombaci, T. T. S. Kuo, and U. Lombardo, Phys. Rep. {\bf 242},
165 (1994).
\bibitem{tma5}
W. Zuo, Z. H. Li, A. Li, and G. C. Lu, Phys. Rev. C {\bf 69},
064001 (2004).
\bibitem{sz1}
S. S. Zhang, L. G. Cao, U. Lombardo and P. Schuck, Phys. Rev. C
{\bf 93} 044329 (2016).
\bibitem{tma6}
P. Wang and W. Zuo, Chin. Phys. C {\bf 39} 014101 (2015).
\bibitem{tma7}
P. Yin, J.-M. Dong, and W. Zuo, Chin. Phys. C {\bf 41} 114102
(2017).
\bibitem{tmar1}
D. Alonso and F. Sammarruca, Phys. Rev. C {\bf 67}, 054301 (2003).
\bibitem{tmar2}
F. Sammarruca, B. Chen, L. Coraggio, N. Itaco, and R. Machleidt,
Phys. Rev. C {\bf 86}, 054317 (2012).
\bibitem{tmar3}
F. Sammarruca, Eur. Phys. J. A {\bf 50}, 22 (2014).
\bibitem{th1}
H. Tong, X.-L. Ren, P. Ring, S.-H. Shen, S.-B. Wang, and J. Meng,
Phys. Rev. C {\bf 98}, 054302 (2018).
\bibitem{ebhf}
W. Zuo, I. Bombaci, and U. Lombardo, Phys. Rev. C {\bf 60} 024605
(1999).
\bibitem{3bf1}
N. Kaiser, Eur. Phys. J. A {\bf 48}, 58 (2012).
\bibitem{3bf2}
A. Dyhdalo, R. J. Furnstahl, K. Hebeler, and I. Tews, Phys. Rev. C
{\bf 94}, 034001 (2016).
\bibitem{em1}
J. Cugnon, A. Lejeune, and P. Grang¡äe, Phys. Rev. C {\bf 35},
R861 (1987).
\bibitem{em2}
G. F. Bertsch, P. F. Bortignon, and R. A. Broglia, Rev. Mod. Phys.
{\bf 55}, 287 (1983).
\bibitem{em3}
D. Page, J. M. Lattimer, M. Prakash, and A. W. Steiner, ApJS {\bf
155}, 623 (2004).
\bibitem{em4}
M. Baldo and G. F. Burgio, Rep. Prog. Phys. {\bf 75}, 026301
(2012).
\bibitem{em5}
X. L. Shang, A. Li, Z. Q. Miao, G. F. Burgio, and H.-J. Schulze,
Phys. Rev. C {\bf 101} 065801 (2020).
\bibitem{em6}
P. Yin, X.-H. Fan, J.-M. Dong, W.-M. Guo and W. Zuo, Nucl. Phys. A
{\bf 961}, 200 (2017).
\bibitem{mr1}
W. Zuo, U. Lombardo, H.-J. Schulze, and Z. H. Li, Phys. Rev. C
{\bf 74} 014317 (2006).













\end{thebibliography}
\end{document}